\documentclass[12pt,preprint]{aastex}
\slugcomment{submitted to {\it The Astronomical Journal}}
\begin{document}
\title{$uvbyCa$H$\beta$ CCD Photometry of Clusters. VII. The Intermediate-Age Anticenter Cluster Melotte 71}
\author{Bruce A. Twarog\altaffilmark{1}, Stuartt Corder\altaffilmark{2}, and 
Barbara J. Anthony-Twarog\altaffilmark{1}}
\affil{Department of Physics and Astronomy, University of Kansas, Lawrence, KS 66045-7582}
\affil{Electronic mail: btwarog@ku.edu, sac@astro.caltech.edu, bjat@ku.edu}
\altaffiltext{1}{Visiting Astronomer, Cerro Tololo Inter-American Observatory.
CTIO is operated by AURA, Inc.\ under contract to the National Science
Foundation.}
\altaffiltext{2}{Current Address:Department of Astronomy, California Institute of Technology, Pasadena, CA 91125}
\begin{abstract}
CCD photometry on the intermediate-band $uvbyCa$H$\beta$ system is presented for the anticenter, intermediate-age open cluster, Melotte 71. Restricting the data to probable single members of the cluster using the color-magnitude diagram and the photometric indices alone generates a sample of 48 F dwarfs on the unevolved main sequence. The average $E(b-y)$ = 0.148 $\pm$0.003 (s.e.m.) or $E(B-V)$ = 0.202 $\pm$0.004 (s.e.m.), where the errors refer to internal errors alone. With this reddening, [Fe/H] is derived from  both $m_1$ and $hk$, using H$\beta$ and $b-y$ as the temperature index, with excellent agreement among the four approaches and a final weighted average of [Fe/H] = --0.17 $\pm$0.02 (s.e.m.) for the cluster, on a scale where the Hyades has [Fe/H] = +0.12. When adjusted for the higher reddening estimate, the previous metallicity estimates from Washington photometry and from spectroscopy are now in agreement with the intermediate-band result. From comparisons to isochrones of appropriate metallicity, the cluster age and distance are determined as 0.9 $\pm$0.1 Gyr and $(m-M)$ = 12.2 $\pm$ 0.1 or $(m-M)_0$ = 11.6 $\pm$ 0.1. At this distance from the sun, Mel 71 has a galactocentric distance of  10.0 kpc on a scale where the sun is 8.5 kpc from the galactic center.  Based upon its age, distance, and elemental abundances, Mel 71 appears to be a less populous analog to NGC 3960.
\end{abstract}

\section{INTRODUCTION}
This is the seventh paper in an extended series detailing the derivation of fundamental parameters in star clusters using precise intermediate-band photometry to identify probable cluster members and to calculate the cluster's reddening, metallicity, distance and age. The initial motivation for this study was provided by \citet{tat97}, who used a homogeneous open cluster sample to identify structure within the galactic abundance gradient. The open clusters appear to populate a bimodal abundance distribution with the clusters interior to a galactocentric distance of $\sim$10 kpc averaging [Fe/H] $\sim$ 0.0 while those beyond this boundary average [Fe/H] $\sim -0.3$. The dispersion about each peak is less than 0.1 dex and implies that within each zone, no statistically significant abundance gradient exists, in contrast with the traditional assumption of a linear gradient over the entire disk. The absence of a simple linear gradient over the extended length of the disk has since been supported through the investigations of Cepheids \citep{an12,an02,lu03}, OB stars \citep{da}, open clusters \citep{at05, di05, yo05}, and field giants \citep{cy05,fu06}.  The exact origin and reason for the survival of the discontinuity remain unresolved issues, though delayed evolution of the outer disk through mergers \citep{tat97,yo05} and the radial variation in the impact of spiral structure on star formation and stellar dynamics appear to be potential contributing factors \citep{sc01,mi02,lp03}. 

Detailed justifications of the program and the observational approach adopted have been given in the earlier papers in the series \citep{AT00a,AT00b,TW03,at04} (hereinafter referred to as Papers I through IV) and will not be repeated. Suffice it to say that the reality of the galactic features under discussion will remain questionable unless the error bars on the data are reduced to a level smaller than the size of the effect being evaluated or the size of the sample is statistically enhanced. The overall goal of this project is to do both and, as discussed in \citet{actt}, a well-defined photometric approach is more than adequate to achieve the goal.

An equally important aspect of this research is detailed testing of stellar evolution theory as exemplified by comparisons to stellar isochrones based upon models derived under a variety of assumptions. The agreement with (or the deviation from) the predicted distribution and location of stars within the CMD has consistently provided a lever for adjusting our degree of confidence in the specifics of stellar interiors as a function of mass and age. A valuable illustration of this can be found in the discussion of the red giant branch distribution (first-ascent and clump giants) of NGC 3680 in comparison to NGC 752 and IC 4651, clusters of comparable age (Paper IV).

The focus of this paper is the supposedly metal-deficient, intermediate-age open cluster, Melotte 71 (hereinafter referred to as Mel 71). The research on this cluster might best be described as marginal. In recent years it has gained some attention due to its anticenter position ($l, b =$ 229\arcdeg, +4.5\arcdeg) and an [Fe/H] between --0.3 and --0.6, coupled with an estimated galactocentric distance placing it between 10 and 11 kpc on a scale where the sun is at 8.5 kpc.  It therefore lies within the potentially critical transition zone between the inner and outer disk. As we will discuss in Sec. 5, most of these estimates are built upon a foundation of marginal or unreliable photometric studies. Despite this, Mel 71 is now regularly included in studies of the properties of the galactic disk cluster population \citep{gr00,ch03,sa04}.

Section 2 contains the details of the $uvbyCa$H$\beta$ CCD observations, their reduction and transformation to the standard system, and a search for photometrically anomalous stars. In Sec. 3 we discuss the CMD and begin the process of identifying the sample of probable cluster members. Sec. 4 contains the derivation of the fundamental cluster parameters of reddening and metallicity. The cluster age and distance are the focus of Sec. 5, along with past estimates of the cluster parameters. Sec. 6 summarizes our conclusions.

\section{The Data}

\subsection{Observations: CCD $uvbyCa$H$\beta$}
The new photometric data for Mel 71 were obtained using the 0.9-meter telescope at Cerro  Tololo Inter-American Observatory (CTIO) of the National Optical Astronomy Observatories.   The detector was a $2048 \times 2048$ Tektronix CCD mounted at the $f/13.5$ focus of  the telescope.  The field size is approximately $13.5'$ on a side.  We obtained frames of  Mel 71 over two runs in February 1998 and January 1999, using our own $3'' \times 3''$  $Ca$ filter, $4'' \times 4''$ $u,v,b,y$ filters from CTIO, and $3'' \times 3''$ H$\beta$ filters borrowed from Kitt Peak National Observatory. No $Ca$ filter was available for the 1998 run. Processing of all frames through bias subtraction, trimming and flat fielding was accomplished at the telescope using  standard IRAF routines.   

Data from three nights in 1998 and six nights in 1999 were incorporated in the analysis of  the Mel 71 photometry.  For each of our filters, with the exception of $Ca$, 8 to 15 frames  were analyzed as detailed below. For $Ca$, the lack of a functional filter in 1998 led to a  reduction in the usual number of frames to 5. Despite the small number of frames, the  internal precision in the final $hk$ indices is more than adequate for our purposes. The total exposure times represented in our analyzed photometry are 6, 30, 21, 30, 22, 22 and 104  minutes for  the $y, b, v, u, Ca$, H$\beta$ wide and narrow filters, respectively.

\subsection{Reduction and Transformation}
Previous papers in this series describe the procedures used to produce high precision,  accurately calibrated photometry from CCD data. IRAF Allstar routines are used to obtain  complete sets of profile-fit magnitudes for all stars on every program frame.  Paper I provides a comprehensive description of the steps used to produce a set of  average indices of high internal precision for the stars in the program cluster. Regardless of the internal precision achieved by averaging large numbers of frames, the  accuracy of the photometric calibration is limited by other factors, including the breadth of parameter space covered by standard stars and observational conditions. Our approach for  these steps is described extensively in Paper IV with a brief delineation presented here for  Mel 71's calibration.

While frames from both 1998 and 1999 runs were used to compile the set of instrumental indices from profile-fit algorithms, the transformation to standard systems depended entirely on several photometric nights from the January 1999 run. Standard stars in the field and in clusters, as well as uncrowded stars in the program cluster, are observed on photometric nights and reduced using corrections for extinction determined for each night and identical (in this case, large, $\sim$16 pixel radius) apertures and sky annuli with comparable areas. Our approach allows for the standardization of aperture photometry in the program cluster obtained on any photometric night. By extension, we are able to apply calibration relations to the high precision profile-fit photometric indices by determining the  mean difference between the calibrated aperture photometry for a photometric night's data  in the program cluster and the profile-fit indices. The precision with which these aperture corrections were determined for H$\beta$ indices in the cluster was 0.003 to 0.004 mag, where these figures reflect a standard error of the mean difference between profile-fit and aperture measurements for 20 to 30 stars.  A larger number of stars in common for the remaining Str\"omgren indices (60 to 100) is reflected in smaller standard errors, 0.002 to 0.003 for the mean aperture correction terms for $V$, $b-y$, $m_1$, $c_1$ and $hk$. 

Our sources for standard stars include datasets in clusters and field star catalogs.  For  uniform $V$ magnitudes, $(b-y)$ colors and $hk$ indices, our primary source is our own  1995 catalog \citep{tat95}. For H$\beta$ values, we used the catalog of \citet{sn88} for six field stars in addition to datasets in several open clusters, including M67 \citep{ntc}, NGC 2287 \citep{glh} and NGC 3766 \citep{sho85,sho87}.  These references also include the $uvby$ photometry used exclusively to calibrate the $m_1$ and $c_1$ indices for main sequence stars.  Whenever possible, we use cluster datasets primarily to establish the slopes and color terms relating instrumental to standard indices and well-observed field stars standards to establish the zero points of calibration equations for each photometric night.  This is not always feasible for some runs with heavier-than-usual reliance on standard star sets from open clusters.

Three photometric nights in 1999 were devoted to H$\beta$ observations in Mel 71 and standard open clusters, with field stars observed on only one night of that run. The relationship between standard and instrumental indices was examined for all three nights to determine a common slope, with a separate zeropoint determined from aperture indices of standard stars on each of the three nights.  Because 40 to 60 standard stars were utilized for each of those calibration equations, the standard deviations of 0.024 to 0.035 about the calibration equations imply standard errors of the mean for the zeropoints of 0.004 to 0.006. A calibration equation for H$\beta$ aperture indices for each night implies a calibration equation for H$\beta$ profile-fit indices with application of the aperture correction, itself carrying a comparable standard error of the mean. The zeropoint for our final calibration equation is a weighted mean of the zeropoints from each of the three photometric nights with a standard error of the mean of 0.008 mag.

Similar procedures are followed to calibrate the other indices.  For $V$, $b-y$ and $hk$, observations of field standards on two photometric nights in 1999 form the basis for the calibration.  For both $V$ and $hk$, slight ($\sim$0.03) color terms were applied.  Following past practice, a separate calibration equation for $b-y$ applicable to dwarfs redder than 0.42 was derived but has not been applied to the fairly early-type stars in this relatively young cluster.  Standard errors of the mean zeropoints for the calibration equations for $V$, $b-y$ and $hk$ are 0.007, 0.008, and 0.015 respectively.  The somewhat larger uncertainty in the zeropoint of the $hk$ calibration is traceable to the relatively small number of Mel 71 $Ca$ frames from which the aperture correction could be determined.  The calibration of the $m_1$ and $c_1$ indices for main sequence stars was based on stars in open clusters on one night of the 1999 run.  In all, 82 stars in M67, NGC 2287 and NGC 3766 contributed to the calibration equations.  The standard errors of the mean for the calibration equation zeropoints are 0.005 and 0.010 respectively for $m_1$ and $c_1$.   

The final photometry for all stars with at least two observations in  $y$ and $b$ is given in Table 1. Identifications and XY coordinates are on the system found  in the WEBDA Cluster Data Base. For those stars not included in the WEBDA database, an identification number has been created beginning with 2000. Average standard errors of the mean with one-sigma ranges for the various indices as a function of $V$ are shown in Figs. 1 and 2.

\subsection{Potential Variables}
The photometry discussed above was obtained to reliably define the cluster in intermediate-band CMD and color-color diagrams so, while over 80 frames have been analyzed, the distribution is rather uneven with filter, ranging from a maximum of 5 for $Ca$ to 15 for $b$. Despite the lack of design for detecting variable stars, it should be possible to identify at least some candidates with modest to large amplitude variations over a range of timescales. 

The obvious criterion for discovering variables is a large photometric scatter in the indices for a star relative to what is expected at a given magnitude. The points that deviate from the well-defined trends in Figs. 1 and 2 should generate a first cut for our sample. However, deviant points could exist in Figs. 1 and 2 for a variety of reasons, most of which have nothing to do with variability. Since we are plotting the standard error of the mean, two stars at the same magnitude level with identical frame-to-frame photometric errors will populate different locations in the figure because one star only appears on a small subset of the frames. The lack of a complete set of magnitudes is a common occurrence for stars in the outer edges of the cluster field, as the cluster position on the CCD chip is shifted from night to night and run to run to avoid placing the same stars on the same bad columns and pixels, and to extend the field coverage beyond the minimum allowed by one CCD frame. Additionally, a paucity of measures for a star in any region of the cluster may be a sign of image overlap/confusion, resulting in poor to inadequate photometry. Finally, while the frames have been processed using a variable point-spread-function, the success of such modifications may decline as one nears the edge of the chip and the photometric scatter in any filter may be expected to increase in the outer zone of the field, irrespective of the number of frames included in the average.

To identify potential variables, we have renormalized the standard errors of the mean to a common number of frames for all stars down to $V$ = 17.  Stars that deviated significantly from the mean error relations at a given magnitude in both $V$ and $b-y$ were selected. These stars were then checked for contamination from nearby stars and eliminated if there was another star within 2 pixels of the supposed variable. Finally, because of the large number of H$\beta$ frames, a secondary check on variability was made using this index and stars were retained only if they exhibited significant scatter for their magnitude in H$\beta$. The final sample consists of eight stars: 114, 218, 2180, 2182, 2205, 2206, 2271, and 2287. These stars are plotted as filled circles in Figs. 1 and 2. The only search for variable stars within Mel 71 to date has been that of \citet{ki99}, who discovered four $\delta\ $ Scuti stars and one eclipsing binary. None of the $\delta\ $ Scuti stars has been tagged by us as variables, not unexpected given the small amplitudes, but the eclipsing variable, V5, discovered by \citet{ki99} is star 218. Six of the candidate variables are outside the field of \citet{ki99}.

\subsection{Comparison to Previous Photometry: Photoelectric Broad-Band $V$}
While there have been two photographic studies of Mel 71, the photometry of \citet{ha76} has been shown to be unreliable with large zero-point errors and low internal precision \citep{po86, me97} while the survey by \citet{po86} is based upon only one photographic plate in each color. We will focus on two sources of photoelectric data, \citet{po86} and \citet{me97}, and the one broad-band CCD survey available to date, \citet{ki99}. 
Data for the published surveys have been taken from the WEBDA Cluster Data Base. 

Single photoelectric $BV$ observations are available for 13 stars ranging from $V$ = 9.43 to 16.68 from \citet{po86}, though no data are provided on the photometric accuracy of the standards obtained with the CTIO 0.9-meter telescope.  All 13 stars are included in our CCD sample, but only 11 were listed with WEBDA identifications. Star M definitely is not included in the WEBDA listing, but star D of \citet{po86} appears to be WEBDA 337 and/or 1031. From 13 stars, the average residual in $V$, in the sense (PJ - Table 1) is  --0.029 $\pm$0.026. If one star (102) with an anomalous residual is removed, the average residual in $V$ becomes --0.033 $\pm$0.018 from 12 stars, testifying to the surprisingly high precision of the photoelectric data, even at fainter magnitudes.

The second photoelectric survey \citep{me97} focuses on the giant branch, supplying $UBV$ data for 22 stars from $V$ = 9.4 to 13.3. Of these, 20 are within our CCD field. The average residual in $V$, in the sense (ME - Table 1), is --0.014 $\pm$0.048. If one star (211) with an anomalous residual is removed, the average residual in $V$ is reduced to --0.006 $\pm$0.032. We note in passing that the $V$ value for star 211 from Table 1 is in excellent agreement with the CCD value of \citet{ki99}, suggesting that the $V$ in \citet{me97} is too bright by 0.17 mag. In neither comparison with the photoelectric surveys is there evidence for a color dependence among the residuals. 

As in the previous papers in the series, we will compare the residuals in the CCD  photometry \citep{ki99} for the entire sample and for the brighter stars alone since the reliability of the latter sample is more relevant to the determination of the cluster parameters. Our interest in the residuals is twofold: to measure the photometric uncertainties in comparison with what is expected from the internal errors and to identify stars that exhibit significant deviations from one study to another. The deviants will be a mixture of misidentifications, bad photometry, and, of particular value, long-term variables or longer-period eclipsing binaries that will not be readily exposed by the variable star searches done to date. 

For 421 stars at all $V$, the residuals, in the sense (Table 1 - K99), average +0.028$\pm$0.198.  While in the mean the systems compare well, the scatter is clearly unacceptable and is dominated by large residuals at fainter magnitudes below 17.0. If we only use stars within 0.10 magnitudes of the mean between the systems, the average residual reduces to +0.007 $\pm$0.047 for 273 stars.  If we restrict the sample to 220 stars brighter than $V$ = 17.0 within 0.2 mag of the mean residual, the average residual becomes --0.001 $\pm$ 0.059. Dropping the 17 stars with a residuals greater than 0.10 mag, the average becomes --0.005 $\pm$0.045, with no evidence for a color term. Finally, limiting the data to the range that overlaps with the photoelectric systems, $V < 15$, from 98 stars, the average residual in $V$ is --0.005 $\pm$0.042. Excluding the 3 stars (40, 62, and 202) with residuals above 0.1 mag, the average residual in $V$ becomes --0.006 $\pm$0.039. In summary, for the magnitude range of interest in this study, the $V$ data of \citet{ki99} appear to be on the same system as Table 1 within 0.01 mag; at fainter magnitudes, the errors in the earlier CCD study grow rapidly. It should be noted that, even at brighter magnitudes, a side-by-side comparison of the $V, b-y$ CMD from Table 1 with the $V,B-V$ CMD of \citet{ki99} for the stars that overlap both studies demonstrates that the dominant source of the photometric scatter lies with the latter reference.

\section{The Color-Magnitude Diagram: Thinning the Herd}
For a program cluster like Mel 71, a key step before deriving the cluster parameters is isolation of probable cluster members with high precision photometric indices. While a location in the galactic anticenter near the plane ($b=$ +4.5\arcdeg) creates a potential issue with field star contamination, the problem is compounded by the modest richness of the cluster and the absence of membership information for any stars except the red giants.  As in past papers in the series, the initial step in maximizing the sample of members has been to restrict the sample to stars with high precision photometry within the cluster core. 

The CMD for all stars with two or more observations in both $b$ and $y$ is shown in Fig. 3. Stars with errors greater than $\pm$0.010 in $b-y$ are presented as crosses. At the bright end, the rich population of giants and the cluster turnoff are readily apparent. The giant branch and red clump are ill-defined, despite the significant number of evolved stars, indicative of an age below 1 Gyr, consistent with the color and luminosity differential between the giants and the main sequence turnoff. The small errors in $b-y$ are maintained consistently down to $V$ = 16.25. The field star contamination becomes significant for $V$ $\sim$17, but the unevolved main sequence is still identifiable amid the CMD distribution down to $V$ $\sim$17.5. 

In an attempt to reduce field star contamination, we next restrict the data to the spatial core of the cluster. We applied the same technique based upon radial star counts used in past papers in this series to identify the probable center of the cluster near star 1143, although the modest richness of the cluster makes this imprecise. The radial profile of the cluster for stars with $V < 18$ is illustrated in Fig. 4. The surface density has been normalized so that the average value in the outermost rings is approximately 1.0. As is apparent, the cluster is not very concentrated and is difficult to identify above the level of the background near R = 100 units, equivalent to about R = 4$\arcmin$ using the WEBDA coordinate scale of 24.2 units/arcmin. For the next cut, stars within 75 coordinate units of the newly defined center were retained. The improvement in the homogeneity of the sample is illustrated by Fig. 5, where the core stars are plotted using the same symbols as in Fig. 3. The delineation of the cluster CMD is greatly enhanced, while the majority of the giant distribution remains intact. The main sequence is well-defined between $V$ = 14 and 17, with evidence for a binary sequence extending 0.75 mag above the probable single-star relation. Note that the ranges for $b-y$ and $V$ are smaller than in Fig. 3.

\subsection{Thinning the Herd: CMD Deviants}
For purposes of optimizing the derivation of the cluster reddening and metallicity, our interest lies in using only single stars that evolve along a traditional evolutionary track and that have indices in a color range where the intrinsic photometric relations are well defined. With this in mind, we next cut the sample to include only stars in the magnitude range from $V$ = 14.5 to 17.0 and the color range from $b-y$ = 0.25 to 0.55, thereby eliminating blue stragglers, some binaries, evolved subgiants, the red giant branch, and fainter stars with larger probable photometric errors. The 420 stars of Fig. 5 are thereby reduced to 122. It should be noted that the CMD limits were chosen with the H$\beta$ photometry in hand to aid in identifying stars within the F-star calibration range; while some stars in this cut of the CMD have H$\beta$ values that indicate they are not F-stars, no star outside this cut has photometric indices that indicate it is an F star. Finally, we make a further restriction that the photometric errors be less than or equal to 0.020 and 0.015 for $m_1$ and H$\beta$, respectively. No restrictions are placed on the other indices. This reduces our sample to 105 stars.

It is highly probable that the majority of stars within this photometric box are members, though not necessarily single stars. As in past papers in the series, it is valuable to identify and remove the likely binary systems from the sample to avoid distortions of the indices through either anomalous evolution or simple photometric combinations of the light for stars in significantly different evolutionary states. Unlike past clusters in this series, Mel 71 is young enough that the F-star range is below the cluster turnoff by at least a magnitude. Thus, all the F stars should exhibit minimal evolution and their CMD distribution should resemble the unevolved main sequence in slope. This is confirmed by the CMD for our 105 stars in Fig. 6. Keeping in mind that the cluster turnoff lies at least 0.5 mag brighter than the limit of this figure, any scatter in the CMD must be caused by field stars, binaries, and/or photometric errors. If we believe our photometry in $b-y$ has the precision we claim, stars that deviate from the well-defined main sequence found in Fig. 6 must fall into one of the categories above, predominantly the first two. Stars that deviate from the main sequence by an unusual degree have been tagged in Fig. 6 as filled circles, though clearly there are some stars that could be classified as borderline members of either group, member or deviant. 

To check the reality of our choices and, indirectly, our claim that the source of the deviation is a real color difference rather than simple photometric scatter, we first use $v-y$, a color index with a larger baseline in wavelength and greater sensitivity to temperature change than $b-y$. Moreover, $v$ is dominated by metallicity effects rather than surface gravity and all the stars within the cluster supposedly have the same [Fe/H].  Fig. 7 shows the $V, v-y$ diagram  for the same 105 stars of Fig. 6 with the same symbol for each. As hoped, the agreement is excellent, with only a pair of stars tagged as deviants falling within the zone dominated by the normal stars. The final test is supplied in Fig. 8 using the $Ca-y$ index, with the largest color baseline of the three plotted. The conclusion remains the same. It is important to recognize the change of scale within the temperature index when discounting the option of photometric errors as the source of the scatter. The range in color among the select stars in the CMD is 0.3 mag, 0.6 mag, and 0.8 mag for $b-y$, $v-y$, and $Ca-y$, respectively. Though there may be a handful of stars that could be added to the list of deviants based upon their positions in Fig. 7 and Fig. 8, we will exclude only those 34 stars already tagged in Fig. 6.

\section{Fundamental Properties: Reddening and Metallicity}
\subsection{Reddening}
With 71 stars selected in a tightly constrained region of the cluster main sequence, we can now calculate individual reddening values, correct for a cluster mean reddening, and use the individual metallicity estimates to define a cluster mean [Fe/H].  As discussed in Paper I, derivation of the reddening from intermediate-band photometry is a straightforward, iterative process given reliable estimates of H$\beta$ for each star. Since metallicity and reddening both affect the intrinsic colors and the evaluation of each parameter, the reddening is derived for a range of assumed values for $\delta$$m_1$($\beta$), the metallicity index, then the metallicity index is derived for a range of assumed reddening values. Only one combination of $E(b-y)$ and $\delta$$m_1$ will simultaneously satisfy both relations. The primary decision is the choice of the standard relation for H$\beta$ versus $b-y$ and the adjustments required to correct for metallicity and evolutionary state. The two most commonly used relations are those of \citet{ols88} and \citet{ni88}. As found in previous papers for IC 4651, NGC 6253, NGC 3680, NGC 2243, and NGC 2420, both produce very similar if not identical results. 

Processing the indices for the 71 stars through both relations generates $E(b-y)$ = 0.146 $\pm$0.025 (s.d.) with \citet{ols88} and $E(b-y)$ = 0.149 $\pm$0.023 (s.d.) with \citet{ni88}, with $\delta$$m_1$($\beta$) = 0.026 $\pm$0.004 (s.e.m.) from 48 stars falling within the H$\beta$ range for F dwarfs. 
The 23 stars excluded are either cluster members that are too blue (A stars), too red (G stars), or likely field stars with reddening-temperature combinations that placed them fortuitously within the band of unevolved main sequence stars, based upon the H$\beta$ indices. 
As a compromise, we will take the average of the two and use $E(b-y)$ = 0.148 $\pm$0.003 (s.e.m.) or $E(B-V)$ = 0.202 $\pm$0.004 (s.e.m.) in the analyses that follow.

\subsection{Metallicity from $m_1$} 
Given the reddening of $E(b-y)$ = 0.148, the derivation of [Fe/H] from the $m_1$ index is as follows. The $m_1$ index for a star is compared to the standard relation at the same color; the difference between them, adjusted for possible evolutionary effects, is a measure of the relative metallicity. Though the comparison of $m_1$ is often done using $b-y$ as the reference color because it is simpler to observe, the preferred reference index is H$\beta$ due to its insensitivity to both reddening and metallicity. Changing the metallicity of a star will shift its position in the $m_1$, $b-y$ diagram diagonally, while moving it solely in the vertical direction in $m_1$, H$\beta$. Moreover, reddening errors do not lead to correlated errors in both $m_1$ and H$\beta$. 

Prior to \citet{at05}, we derived the metallicity using $b-y$ and H$\beta$ as the defining temperature index for $m_1$ with, on average, no statistically significant difference in the outcome.  After the publication of Paper IV, alternative [Fe/H] calibrations based upon $b-y$ and $m_1$ were derived by \citet{no04} for F stars and cooler, calibrations that make use of the reddening-corrected indices rather than differentials compared to a standard relation. This approach was first used successfully by \citet{sn89}, but the primary focus of their work was on metal-deficient dwarfs and concerns about the application of the function to solar-metallicity dwarfs limited its adoption for disk stars. These concerns proved valid for the metallicity calibration for cooler dwarfs where [Fe/H] was systematically underestimated at the metal-rich end of the scale \citep{tw02}. The more extensive recalibrations for F dwarfs and cooler by \citet{no04} are readily applicable to solar and higher metallicity dwarfs at all colors and eliminate the concerns regarding the original functions of \citet{sn89}.  We will derive [Fe/H] without reference to H$\beta$ using the \citet{no04} relation as well as from $\delta$$m_1$(H$\beta$), with the latter supplying an independent check of the calibration relations.

After correcting the indices for the effect of $E(b-y)$ = 0.148, the mean [Fe/H] using the F-star relation of \citet{no04} is  $-0.188 \pm 0.044$  (s.e.m.). In contrast, deriving the differential in $m_1$ relative to the standard relation at the observed H$\beta$,  the average $\delta$$m_1$ for 48 stars definable as F stars is $0.026 \pm 0.004$ (s.e.m.), which translates into [Fe/H] $= -0.172 \pm 0.044$ (s.e.m.) for the calibration as defined in \citet{ni88} and adopted in previous papers.  
We note that the zero-point of the H$\beta$ metallicity calibration has been fixed to match the adopted value for the Hyades of [Fe/H] =$ +0.12$, i.e., if one processes the data for the Hyades or the standard relation through the [Fe/H] calibration, one is guaranteed to obtain [Fe/H] = +0.12 for any star with $\delta$$m_1$ = 0.000. 

In contrast, we have in past papers applied an adjustment to the metallicity derived from the \citet{no04} relation.  
If the standard relation or the observed data for the Hyades are processed through the \citet{no04} relation, at the cooler end of the scale beyond $b-y$ = 0.32, one obtains [Fe/H] between = +0.12 and +0.16. As $b-y$ decreases, [Fe/H] declines steadily, reaching a minimum near +0.03 near the hotter end of the scale ($b-y$ = 0.23). Because the stars in the color range of interest in Mel 71 cover the entire F-star range, the mean [Fe/H] should be on a system where the Hyades has the same approximate value as adopted for the H$\beta$ calibration and no adjustment to the \citet{no04} metallicity has been applied.  

The primary weakness of metallicity determination with intermediate-band filters is the sensitivity of [Fe/H] to small changes in $m_1$; the typical slope of the [Fe/H], $\delta$$m_1$ relation is 12.5. Even with highly reliable photometry, e.g., $m_1$ accurate to $\pm$0.015 for a faint star, the uncertainty in [Fe/H] for an individual star is $\pm$0.19 dex from the scatter in $m_1$ alone. When potential photometric scatter in H$\beta$ and $b-y$ are included, errors at the level of $\pm$0.25 dex are common, becoming even larger for polynomial functions of the type discussed above. As noted in previous papers in this series, the success of the adopted technique depends upon both high internal accuracy and a large enough sample to bring the standard error of the mean for a cluster down to statistically useful levels, i.e., below $\pm$0.10 dex. Likewise, because of the size of the sample, we can also minimize the impact of individual points such as binaries and/or the remaining nonmembers, though they will clearly add to the dispersion.

\subsection{Metallicity from $hk$}
We now turn to the alternative avenue for metallicity estimation, the $hk$ index. The $hk$ index is based upon the addition of the $Ca$ filter to the traditional Str\"{o}mgren filter set, where the $Ca$ filter is designed to measure the bandpass that includes the H and K lines of Ca II. The design and development of the $Caby$ system have been laid out in a 
series of papers discussing the primary standards \citep{att91}, an extensive catalog of field star observations \citep{tat95}, and calibrations for both red giants \citep{att98} and metal-deficient dwarfs \citep{at00}. Though the system was optimally designed to work on metal-poor stars and most of its applications have focused on these stars \citep{atc95,bd96,at00}, early indications that the system retained its metallicity sensitivity for metal-rich F dwarfs have been confirmed by observation of the Hyades and analysis of nearby field stars \citep{at02}. What makes the $hk$ index, defined as $(Ca-b)-(b-y)$, so useful for dwarfs, even at the metal-rich end of the scale, is that it has half the sensitivity of $m_1$ to reddening and approximately twice the sensitivity to metallicity changes. The metallicity calibration for F stars derived in \citet{at02} used $\delta hk$ defined relative to $b-y$ as the temperature index. To minimize the impact of reddening on metallicity, this calibration was redone in Paper III using H$\beta$ as the primary temperature index, leading to the preliminary relation
\medskip
\centerline{[Fe/H]$ = -3.51 \delta hk(\beta) + 0.12$}
\smallskip
with a dispersion of only $\pm$0.09 dex about the mean relation. Though the derived zero-point of the relation was found to be +0.07, it was adjusted to guarantee a Hyades value of +0.12, the same zero-point used for the $m_1$ calibration. Because of the expanded sample of stars with spectroscopic abundances and the revised $m_1$ calibrations for F and G dwarfs by \citet{no04}, a revised and expanded calibration of the $\delta$$hk$ indices, based on both $b-y$ and H$\beta$ is underway.  Modest changes have been generated in the color dependence of the [Fe/H] slope for $\delta$$hk(b-y)$, with even smaller adjustments to the H$\beta$-dependent relation. To ensure that that the metallicities based upon $m_1$ and $hk$ are on the same internal system, we have calculated [Fe/H] from the unmodified $m_1, b-y$ function of \citet{no04} for all dwarfs with $hk$ indices and derived linear relations between [Fe/H] and $\delta$$hk(b-y)$ and $\delta$$hk($H$\beta)$ for three different color ranges among the F-stars. 

Applying these modified metallicity calibrations to the $hk$ data for 48 F stars in Mel 71, the resulting [Fe/H] values for $hk$ relative to $b-y$ and H$\beta$ are [Fe/H] = --0.170 $\pm$0.034 (s.e.m.) and --0.150 $\pm$0.031 (s.e.m.), respectively. The fact that the standard errors of the mean for the [Fe/H] values from $hk$ are smaller than the errors from using $m_1$ despite the availability of only 5 CCD frames in $Ca$ is a reflection of the value of coupling the increased metallicity sensitivity of the $hk$ index with the reddening and metallicity-independent H$\beta$ index and the minimal temperature-dependence of the [Fe/H] calibration based upon H$\beta$. The average [Fe/H], weighted or unweighted,  of the four determinations is [Fe/H] = --0.17 $\pm$0.02 (s.d.). 

\section{Discussion}

\subsection{Age and Distance}
With the reddening and metallicity well determined, it is now possible to derive the cluster distance and age through comparison with appropriate isochrones. In past papers, the CMD was constructed using either published broadband $BV$ photometry or, when necessary, by transforming the more precise $b-y$ indices to the $B-V$ system. The latter option was employed because isochrones on the $uvby$ system were unavailable. We will do the same in this investigation because the $b-y$ data for Mel 71 are superior to the $BV$ CCD data but, we will also do a direct test using the $uvby$ Victoria-Regina isochrones \citep{vb06} generated with the transformations of \citet{cl04}. 

To allow the match to the broad-band isochrones, the $b-y$ CCD data has been correlated with the combined photoelectric data of \citet{me97} and the CCD data of \citet{ki99} brighter than $V$ = 15.0. The latter restriction is made to minimize the impact of the increasing scatter in the $BV$ data of \citet{ki99} at fainter magnitudes. Comparison between the two samples indicates that they are on the same $B-V$ system to within 0.005 mag. From 105 stars that overlap in $b-y$ and $B-V$, the following transformations have been found
\medskip
\centerline{$B-V$ = 1.59*$(b-y)$  - 0.019   for $b-y$ $\leq$ 0.70}
\medskip
\centerline{$B-V$ = 1.49*$(b-y)$  + 0.048   for $b-y > 0.70$}
\medskip
with a residual scatter about the mean relation of 0.055 mag.

To minimize issues caused by the adoption of different isochrones for age and distance derivation, we first zero the isochrones of \citet{vb06} by requiring that a star of solar mass at an age of 4.6 Gyr have $M_V$ = 4.84 and $B-V$ = 0.65, the standard adopted in previous cluster analyses in our work. A check of the solar isochrones leads to minor adjustments, $\Delta$$V$ = 0.02 and $\Delta$$B-V$ = +0.013 mag; we assume as the simplest approximation that this offset applies to the isochrone set closest in abundance to the derived cluster metallicity.

Fig. 9 shows the cluster CMD from Fig. 5 superposed upon the [Fe/H] = --0.19, scaled-solar isochrones with ages 0.8, 0.9, and 1.0 Gyr, adjusted for $E(B-V)$ = 0.202 and $(m-M)$ = 12.2. For the giant region of the CMD, the radial-velocity data of \citet{me97} has been used to identify members that are single stars (filled circles), binaries (filled triangles), possible member, single stars (open square), and one possible member, binary star (open triangle). All definite radial-velocity non-members have been removed. Additionally, five of the giant stars within these classes are located outside the core region but are included here to aid in delineating the post-main-sequence evolution. One giant star not included in the $uvby$ survey has been added to the CMD using the $BV$ photometry of \citet{me97}. 

From the comparison between the theory and observation, the match to the unevolved main sequence is reasonable down to the field star confusion limit of the data near $V$ = 17.5 and indicates an age of 0.9 Gyr. The stellar main sequence exhibits slightly more curvature than the isochrones, a point we will return to below. The isochrone hook near the hydrogen exhaustion phase at $V$ $\sim$13.2 appears to extend further to the red than the data and the data appear to curve closer to the 1.0 Gyr isochrone rather than faithfully following the 0.9 Gyr track. It should be remembered that this phase of evolution is rapid, so that a mildly populated cluster like Mel 71 is unlikely to have stars populating the entire track of this isochrone. Moreover, the red hook crosses the region populated by the binary sequence created by less evolved stars between $B-V$ = 0.40 and 0.50, so the exact location of the red hook is subject to significant distortion that cannot be removed without radial-velocity data for the turnoff stars, as exemplified by the detailed analysis of a slightly older cluster, NGC 752 \citep{da94}. 

In the region of the giant branch, the match between theory and observation is less definitive since the isochrones only include the first ascent of the red giant branch and, as will be shown below, the majority of the stars in the giant region are likely to be core He-burning stars. At minimum, the giant branch appears to have two distinct concentrations separated by about 0.5 mag that could be classified as the clump. The majority of the member stars in the lower concentration are binaries, but two of the four stars for which no radial-velocity data are available populate this zone. This bifurcation of the giant branch in clusters of an age comparable to Mel 71, such as NGC 5822 \citep{tam} has been noted before, but its origin remains elusive \citep{gi00, at04}.
Moreover, the fact that the subgiant branch of the isochrones supplies a respectable lower bound to the fainter clump could be a significant test of the quality of the isochrones if one could assume that the single-star luminosity of the clump stars was only mildly impacted by the presence of a companion. With stars on the giant branch, binaries composed of stars with a mass ratio near one can have luminosity ratios larger by one to two orders of magnitude, unlike pairs composed of only main sequence stars. Thus, we have no way of knowing if adjustments to the binaries will place the star significantly fainter than the isochrone.  

We now turn to the isochrones based upon intermediate-band photometry. In this instance, we have no predefined solar color in $b-y$, so we use the models as they are calibrated. Applying a reddening of $E(b-y)$ = 0.148 and $(m-M)$ = 12.2 to the same isochrones as in Fig. 9, we get the comparison of Fig. 10, where the symbols have the same meaning as in Fig. 9. A number of points are immediately apparent. First, a distance modulus slightly larger than in the broad-band fit might supply a closer match between the data and the isochrones, indicating a potential offset of $\sim 0.01$ mag in the adopted $b-y$ scale relative to $B-V$. Second, the main sequence produces a better match to the isochrones in that the isochrones have the same degree of curvature as the data, implying that part of the problem with the fit in Fig. 9 may originate with the simple linear transformation between $b-y$ and $B-V$. Without better broad-band photometry to compare to, this question can't be resolved. Third, the isochrones in the giant region are an excellent match in color and luminosity to the mean location of the stars, keeping in mind that only the first-ascent giant branch is plotted. It is apparent that the 0.9 Gyr isochrone offers the best estimate for the age, which is well bracketed by the 0.8 to 1.0 Gyr range.

For consistency with past analyses, we also refer to the scaled-solar isochrones of \citet{gi02} (hereinafter referred to as PAD). Based upon the many comparisons, including our own \citep{at91, da94, ash, tah}, between open clusters and past and present generations of isochrones, we will only make use of isochrones that include convective overshoot mixing. On a scale where solar metallicity is Z = 0.019 and Y = 0.273, PAD isochrones were also obtained for (Y, Z) = (0.250, 0.008) or [Fe/H] = --0.38. Since PAD isochrones with [Fe/H] = --0.17 are not available but we have a reliable cluster age from our earlier comparison, the cluster data will be required to sit between isochrones with an age of 0.9 Gyr, but [Fe/H] = 0.0 and --0.38. As above, the solar metallicity isochrones were adjusted to match the solar parameters noted earlier, requiring offsets of --0.032 mag in $B-V$ and +0.02 mag in $V$ \citep{tab}. Based upon the analysis of \citet{tab}, these offsets do not appear necessary for the more metal-poor models.

As with the previous comparison, the cluster data do a reasonable job of falling approximately midway between the two metallicities in Fig. 11, though the data exhibit more curvature than predicted by the isochrones. Of particular importance, however, is the morphology of the giant branch. The giants superpose on the more metal-poor isochrone rather than sitting between them, but this is due to the giant branches being too red rather than an indication of a lower [Fe/H]. An offset of $\sim 0.1$ mag is common for the PAD isochrones in this age and metallicity range (Paper IV). What is important is that the luminosity range and distribution of the single stars are consistent with the dual clump being the signature of the vertical band in the post-He-core flash evolution from $V$ = 13.4 to 12.5. It is probable that only one or two stars near the red edge of the giant branch distribution can be first-ascent giants while the majority of the observed stars are evolving from the lower clump to the upper clump, en route to continued evolution on the asymptotic giant branch (AGB). The predicted first-ascent giant branch ceases near $V$ = 11.4, so the brightest members must be AGB stars, assuming the isochrones are a good match to reality. A survey of Li among the giants may confirm if this interpretation is correct (Paper IV).

Thus, $(m-M)$ = 12.2 $\pm$0.1 from both sets of broad-band isochrones. The agreement is effectively hard-wired into the analysis by demanding that the solar isochrones reproduce the same values of the solar color and absolute magnitude. The disagreement with the intermediate-band models is not regarded as a serious problem since it undoubtedly reflects the shortcomings of the transformation process from the theoretical to the observation plane. It is encouraging, however, that both sets of isochrones are consistent with an age of 0.9 $\pm$ 0.1 Gyr for the cluster. After correcting for reddening and the slight difference in the assumed [Fe/H] of the isochrones, $(m-M)_0$ = 11.60 $\pm$0.1 and the cluster distance from the sun is 2090 pc, placing it 163 pc above the galactic plane. 

\subsection{Previous Determinations: Reddening and Metallicity}
The first attempt at a reddening estimate by \citet{ha76} led to $E(B-V)$ = 0.00, though the photographic $UBV$ photometry has since been shown to be highly unreliable. This flawed photographic survey and $E(B-V)$ estimate served as the basis for the first metallicity determination using photographic $\delta$$(U-B)$ from 11 giants by \citet{ja79}, who calculated [Fe/H] = --0.37. The first legitimate attempt at a reddening determination came from \citet{po86} using a comparison of their photographic CMD to the Hyades and requiring alignment of the giant branch clump to get $E(B-V)$ = 0.10 $\pm$0.05. This result was flawed in two ways. First, the only metallicity estimate available implied that Mel 71 was approximately 0.5 dex more metal-deficient than the Hyades and, second, the scattered distribution of giants in the CMD caused by the low precision, photographic data from one plate in each color, non-members, binaries, and the convoluted evolutionary track for Hyades-age giants made the match between the clusters an exercise in optimism. There is little question that both the absolute value and the estimated uncertainty in the reddening were too low.

The low $E(B-V)$ propagates through the next two key studies of the cluster. \citet{ge92} used Washington photometry of the giant branch, coupled with the unpublished radial-velocity data of \citet{me97} to identify field stars and binaries, arriving at [Fe/H] = --0.57 $\pm$0.05 (s.e.m.) from 14 giants. The critical issue for this result is the high sensitivity of the Washington [Fe/H] to small changes in $E(B-V)$. A shift of only 0.05 in $E(B-V)$ increases the metallicity from --0.57 to --0.35. If we double this effect to account for a rise from $E(B-V)$ of 0.10 to 0.20, the new [Fe/H] from Washington photometry becomes --0.13, in excellent agreement with the current value, though some caution should be exercised until a comparison is made between the overall metallicity scale of the Washington system and that of \citet{tat97}.

The only spectroscopic study of the cluster by \citet{br96} used 2 giants to obtain [Fe/H] = --0.32 $\pm$0.16, adopting $E(B-V)$ = 0.10. Though the analysis does not supply a discussion of the impact of changing $E(B-V)$ on the giants in Mel 71, from the related analysis of Tombaugh 2, it is estimated that increasing $E(B-V)$ by 0.1 will raise [Fe/H] by +0.1 dex. Applied to Mel 71, this changes the metallicity to [Fe/H] = --0.22 $\pm$0.16, clearly the same as the photometric value given the large error bars.

The final analysis of the cluster is that of \citet{ki99}, where we find the first hint of a significant reddening value based upon matches between the cluster and theoretical isochrones in $UBV$ and $UVI$ two-color diagrams. As is the case in other papers in this series, it is argued that the large number of stars offsets the large scatter to some extent. A weakness of this approach is the assumed [Fe/H]. Because of the prior papers discussed above, \citet{ki99} adopted a two-color relation based upon isochrones with [Fe/H] = --0.40, significantly more metal-poor than derived here and therefore bluer than expected. Fortunately, the stars at the turnoff are hot enough that metallicity effects on the location of the two-color relations at the turnoff should be smaller than the scatter in the photometric data and the uncertainty in the colors of the isochrones transformed to the observational plane.

Another direct measure of the reddening in the direction of Mel 71 is supplied by the reddening maps of \citet{sc98}, which indicate $E(B-V)$ = 0.26, an upper limit along this line of sight. At the galactic latitude and distance of the cluster, Mel 71 lies 163 pc above the galactic plane. This places it above the majority of the expected vertical distribution of gas and dust that defines the interstellar reddening in the plane. Though hardly definitive proof, the derived $E(B-V)$ of 0.20 is consistent within the uncertainties with the reddening expected for the line of sight of Mel 71 relative to the sun and the center of the galactic plane. A value of $E(B-V)$ = 0.1 would imply that 60$\%$ of the dust lies more than 163 pc above the disk.   

With a reliable age and metallicity for the cluster, we can place one last constraint on the reddening by comparison to other clusters of comparable class. With NGC 5822, one has a slightly more metal-rich cluster, [Fe/H] = ---0.03 \citep{tat97}. The CMD for the core of the cluster bears a striking resemblance to Mel 71 (see Fig. 5 of \citet{tam}).  In particular, the giant branch exhibits the same double clump structure as Mel 71 and the typical color of the vertical band is $B-V$ = 1.01 $\pm$0.02. With the derived cluster reddening of $E(B-V)$ = 0.15 $\pm$0.02, the intrinsic color for the vertical giant branch is $(B-V)_0$ = 0.86. For Mel 71, all things being equal, the comparable clump band is at $B-V$ = 1.04, leading to $E(B-V)$ = 0.18 $\pm$0.03. Since NGC 5822 is slightly metal-rich compared to Mel 71, this should be a lower bound on the reddening.

\section{Summary}
CCD photometry of Mel 71 on the $uvbyCa$H$\beta$ system has been used to isolate a highly probable set of 48 unevolved F-star, cluster members. Analyses of this exclusive sample generates a reddening estimate of $E(b-y)$ = 0.148 $\pm$0.003 (s.e.m.) or $E(B-V)$ = 0.202 $\pm$0.004 (s.e.m.). From both $m_1$ and $hk$ photometry, the cluster has a well-defined [Fe/H] of --0.17 $\pm$0.02. Coupled with scaled-solar isochrones of comparable [Fe/H], the age and distance of the cluster are found to be 0.9 $\pm$0.1 Gyr and $(m-M)$ = 12.2 $\pm$0.1. A close look at these parameters makes Mel 71 a less populous analog to NGC 3960, studied in detail most recently by \citet{br06}. NGC 3960 is found to have an age of 0.9 Gyr from overshoot isochrones, a true distance modulus of $(m-M)_0$ = 11.6, and a spectroscopic metallicity of [Fe/H] = --0.12, in excellent agreement with the value of --0.17 from the composite DDO and moderate-dispersion spectroscopic system of \citet{tat97}, i.e., essentially the same within the errors at Mel 71. Even more intriguing, if one applies corrections to the elemental abundances derived by \citet{br96} to account for the increased reddening, Mel 71 has the same approximate modest enhancement in $\alpha$-elements and large enhancement in Na found among the stars in NGC 3960. If this pattern survives a reduction in the significant error bars on the data in \citet{br96}, as noted by \citet{br06}, NGC 3960 and Mel 71 both fall in line with pattern for the old open cluster population \citep{fr03, ca05}. 

As a single cluster among hundreds, Mel 71 can hardly be called critical to our understanding of the evolution of the disk. It does, however, serve as an example of the quality of information available for the vast majority of clusters commonly included in large-scale analyses of the disk. The photographic photometry of \citet{po86} is of modest accuracy, but supplies a CMD with enough morphological detail to classify the cluster as intermediate-age; more exact estimates have been supplied later by quantitative techniques as exemplified by \citet{sa04}. 
However, the reliability of basic parameters for the cluster continued to be vulnerable to the inevitable
coupling between reddening and metallicity. 
In each of the early photometric studies of Mel 71, the derivation of one parameter was tied to an {\it assumed} value for the other, leading to the circumstance where a reddening estimate tied to Hyades metallicity is then used to derive an abundance close to 1/3 solar which, in turn, is used to estimate a reddening value double the original estimate.  The value of the extended, intermediate-band approach is that it supplies both reddening and metallicity with a minimal level of dependency, an often ignored benefit when selecting observational approaches to cluster studies.

Mel 71 has been observed within the broad-band UBV system, the Washington system, and now the extended Str\"{o}mgren system, including both $m_1$ and $hk$. Observations of the giants on the DDO system would allow a more direct tie-in to the metallicity scale of \citet{tat97}, but the links via the Str\"omgren data should suffice for including it within a discussion of the galactic disk gradient and, in an ideal world, all these determinations should agree. As we have seen with Mel 71, often times they do not or, if they do, the agreement is an illusion coupled with large error bars. One of the more surprising discrepancies is with the results of the Washington photometry \citep{ge92}. The system would seem to be an ideal choice for open cluster studies. It is based upon broad-band filters and focuses on the brightest stars within a cluster, the giants. With a 1.5m telescope and an efficient CCD, one can even observe giants in the Magellanic Clouds with reasonable internal precision \citep{ge87}, using the large statistical sample to lower the standard errors of the mean for the cluster metallicity to practical values. There is no question that the photometry found in \citet{ge92} is high quality and reliable. Why, then, was such a low value derived for [Fe/H] and why does the system occasionally generate even larger discrepancies with other photometric and spectroscopic results, as illustrated by a number of clusters in \citet{ge92}? 

As demonstrated by Mel 71, exquisite internal precision is wasted if the systematic external errors are significant. In this case, as in every case where the reddening estimate has even modest uncertainty, the final error bars in [Fe/H] derived from Washington photometry are dominated by the errors in $E(B-V)$. The key factor is the ratio of $\delta$[Fe/H] to $\delta$$E(B-V)$ For [Fe/H] derived using $m_1$ and $hk$, the ratios are $\sim$3.0 and $\sim$0.5 or less, respectively. For the Washington system, this ratio is $\sim$4.5. Thus, a systematic error of 0.10 in $E(B-V)$ leads to a systematic shift in [Fe/H] of 0.44 dex, as in Mel 71, for the Washington system; a comparable error in $E(B-V)$ systematically shifts [Fe/H] derived from $hk$ photometry by 0.05 dex.

To close, we reiterate a point made in \citet{tat97}. A surprising aspect of the history of the derivation of Mel 71's basic parameters is the minimal impact on its estimated galactocentric distance.  The apparent modulus has consistently been derived as $(m-M)$ = 12.2; the true distance has varied by 15$\%$ solely because of the change in the assumed reddening from $E(B-V)$ = 0.1 to 0.2. The importance of this simple fact is that even if the cluster were in the exact direction of the galactic anticenter, this range in true distance from the sun would translate into a range of less than 3$\%$ in $R_{GC}$, on a scale where the sun is at $R_{GC}$ = 8.5 kpc. Thus, defining the location of Mel 71 within the large-scale picture of the galactic disk  and its abundance gradient is controlled solely by the reliability of its metallicity determination. At $R_{GC}$ = 10.0 kpc and [Fe/H] = --0.17, Mel 71 can be viewed as supporting either picture for the disk,  a two-phase model with approximately flat gradients on either side of a transition near $R_{GC}$ = 10 kpc or a linear gradient of --0.07 dex/kpc \citep{fj02}, with [Fe/H] = 0.0 at $R_{GC}$ = 8.5 kpc. Note that the same conclusion would have been true if [Fe/H] = --0.3 had been derived for the cluster.

\acknowledgements
The progress in this project would not have been possible without the time made available by the NOAO TAC and the invariably excellent support provided by the staff at CTIO. Extensive use was made of the SIMBAD database, operating at CDS, Strasbourg, France  and the WEBDA database maintained at the University of Vienna, Austria (http://www.univie.ac.at/webda/).  The cluster project has been helped by support supplied through the General Research Fund of the University of Kansas and from the Department of Physics and Astronomy.

\clearpage

\figcaption[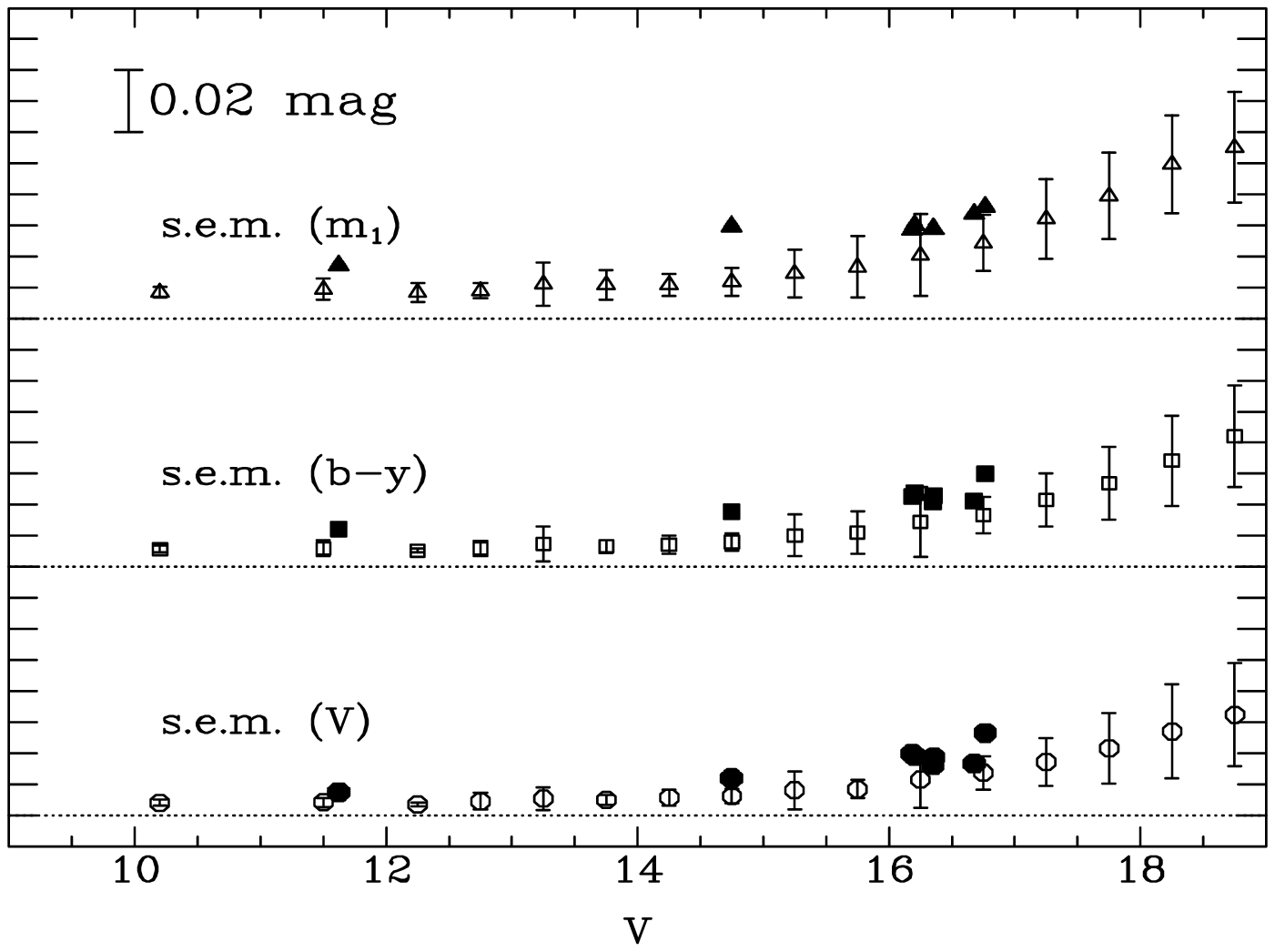]{Standard errors of the mean (sem) for $V$, $b-y$, and $m_1$ as a function of $V$. Open symbols are the average sem while the error bars denote the one sigma dispersion. Filled symbols are stars identified as potential variables. \label{f1}} 

\figcaption[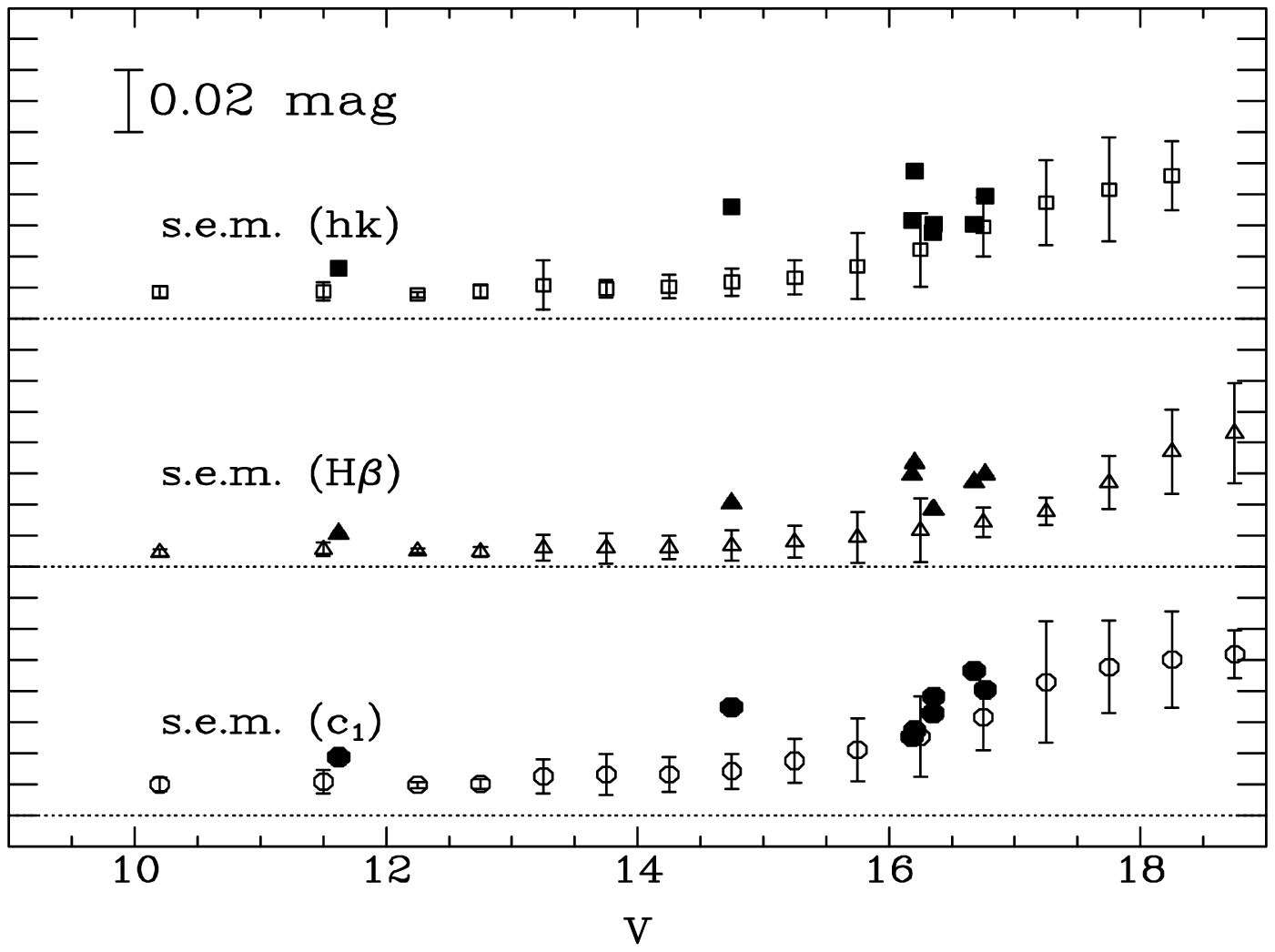]{Same as Fig. 1 for $c_1$, $hk$, and H$\beta$. \label{f2}}

\figcaption[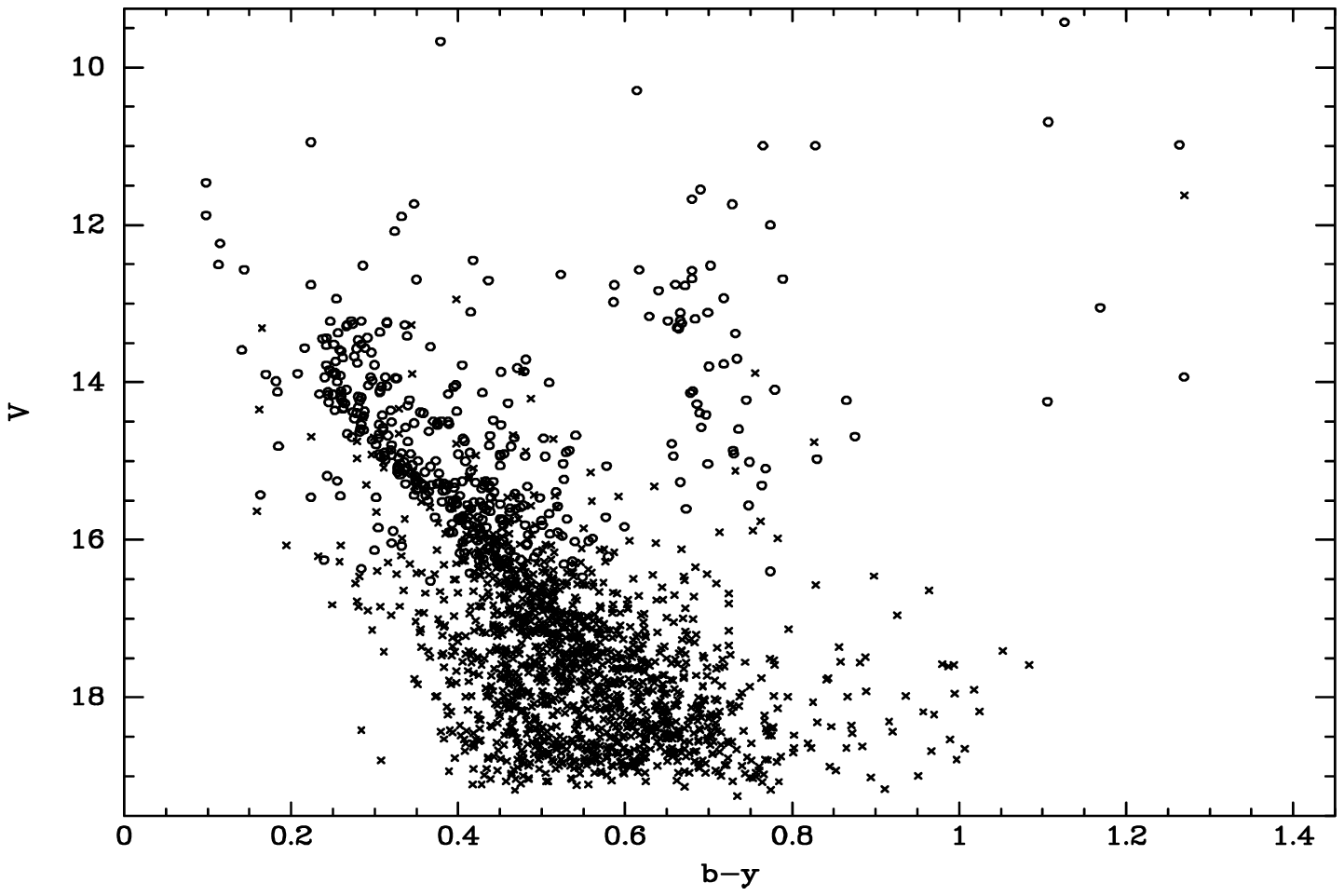]{Color-magnitude diagram for stars with at least 2 observations each in $b$ and $y$. Crosses are stars with internal errors in $b-y$ greater than 0.010 mag.\label{f3}}

\figcaption[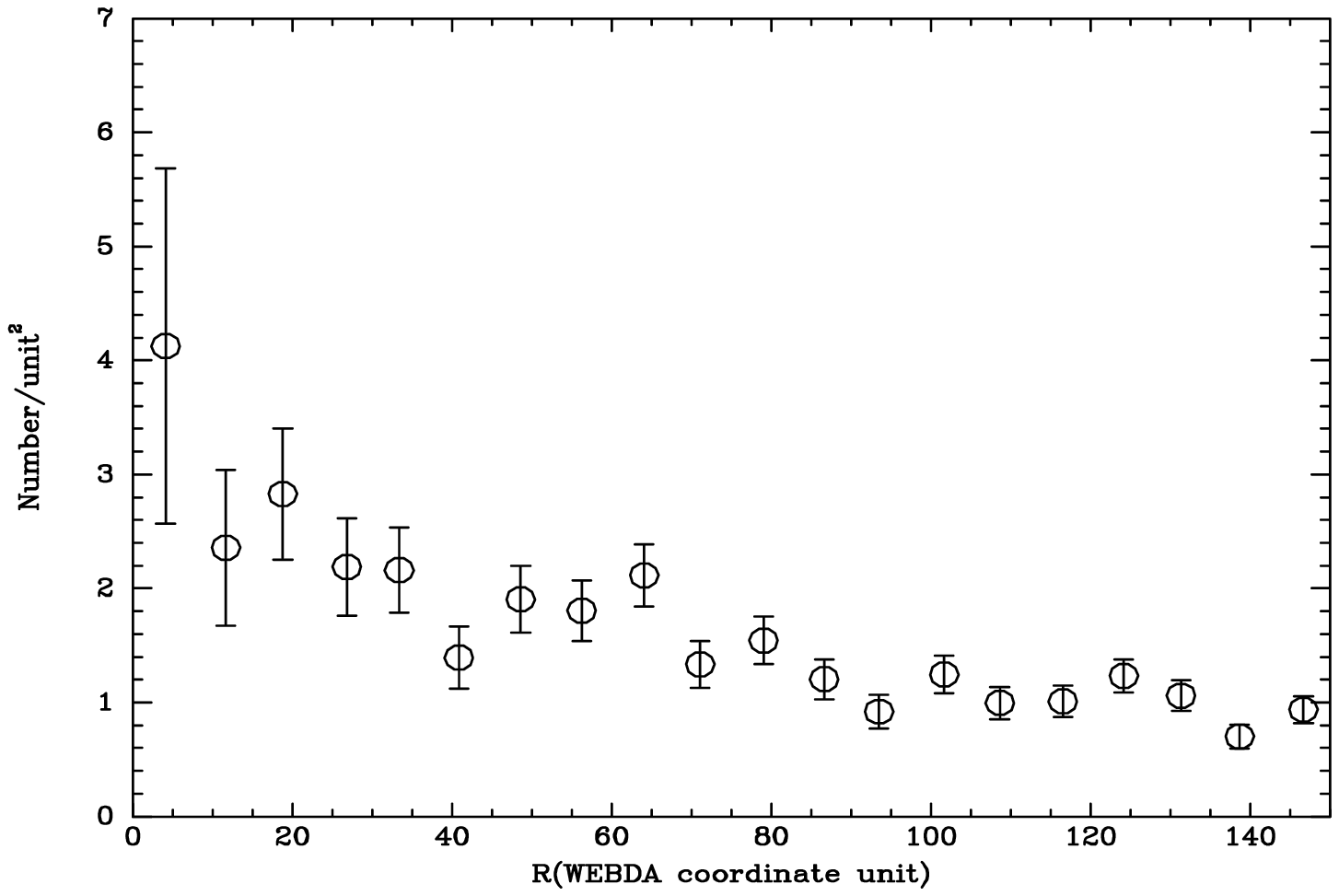]{Stellar surface density as a function of radius in WEBDA coordinate units from the center of the cluster. Density is normalized to 1.0 at the background level of the field.\label{f4}}

\figcaption[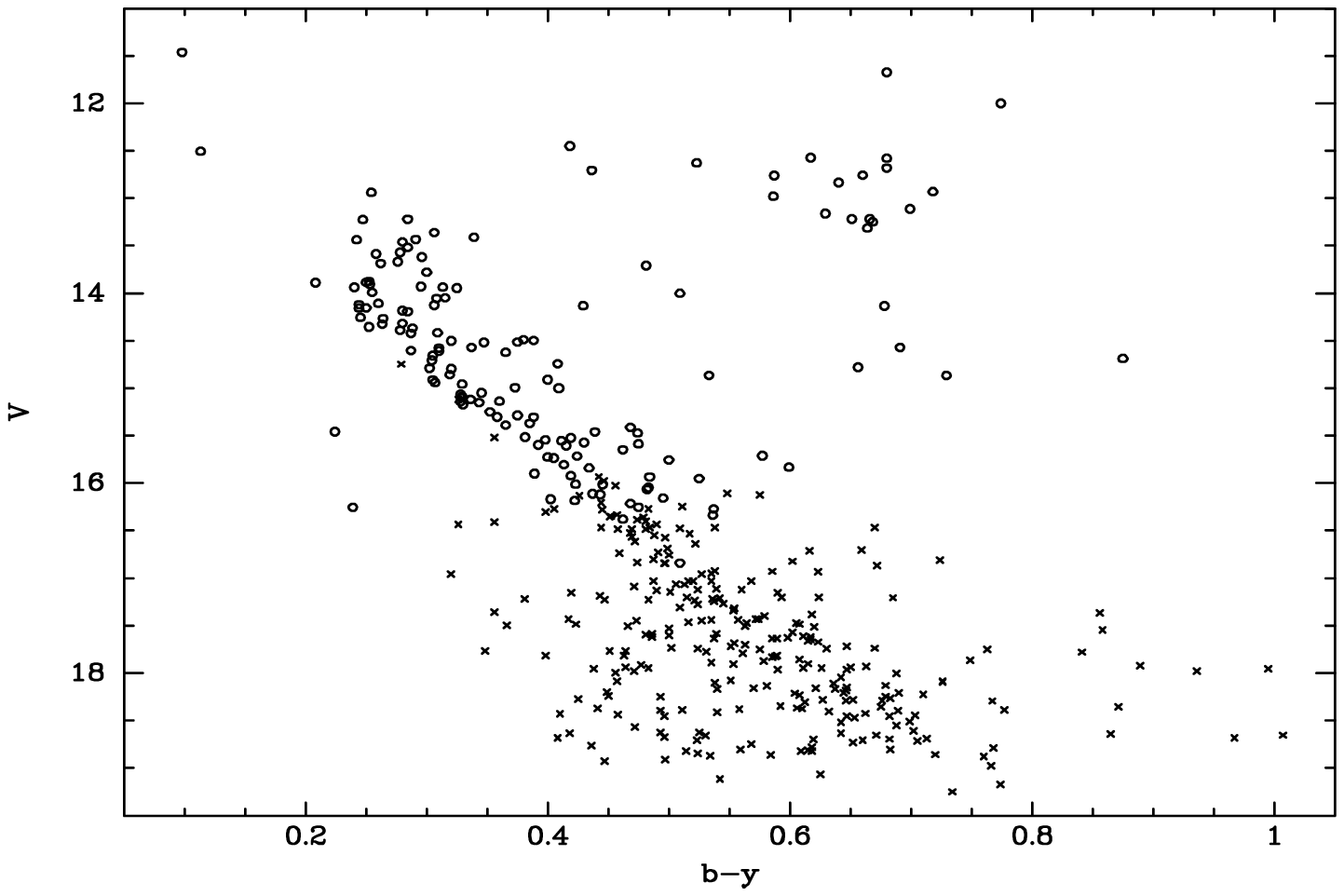] {Same as Fig. 3 for stars within 75 WEBDA coordinate units of the cluster center. \label{f5}}

\figcaption[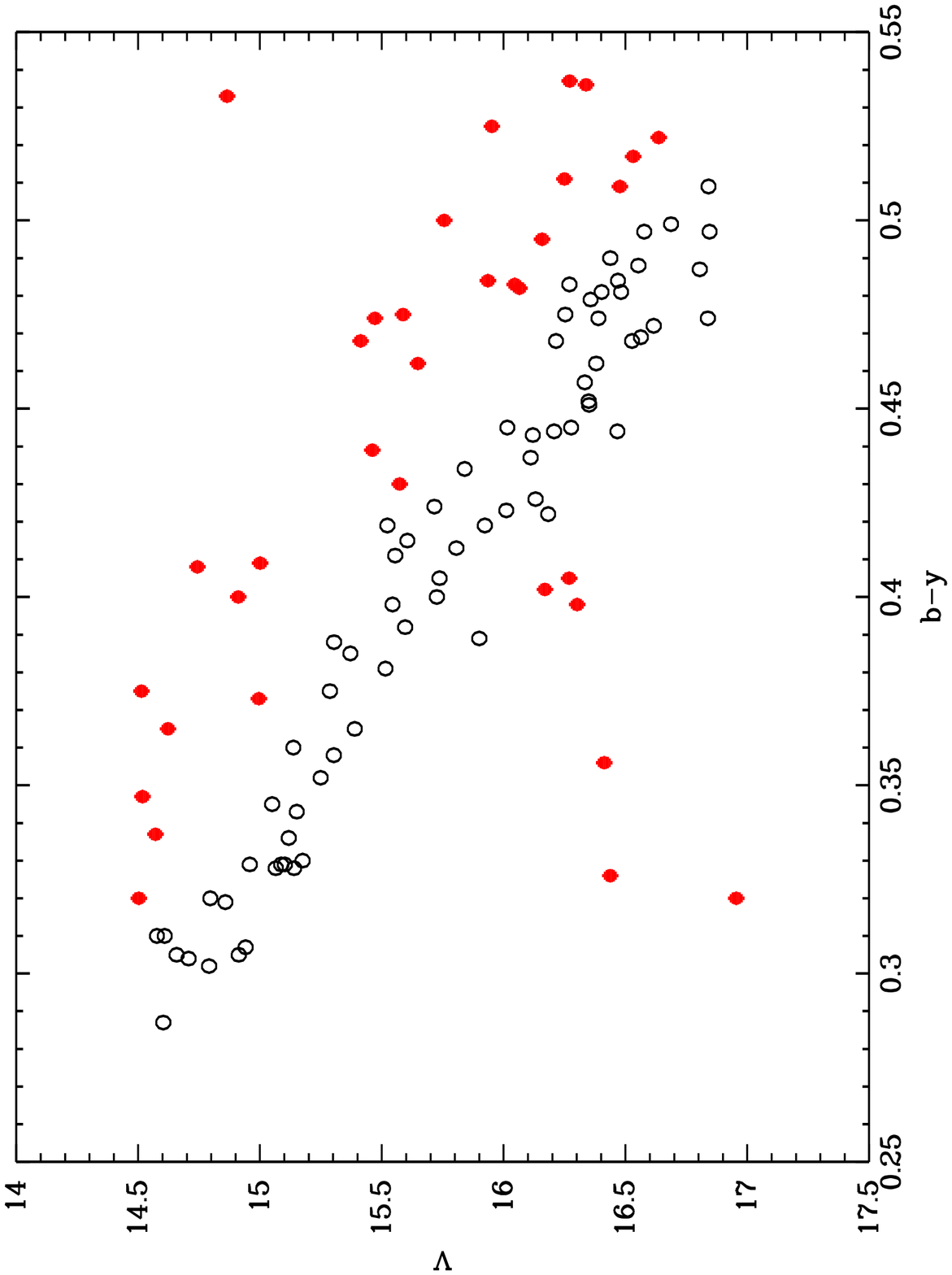]{$V, b-y$ CMD for cluster core stars within the F-dwarf range on the unevolved main sequence. Filled red circles are stars tagged as potential binaries, nonmembers, or photometric anomalies.\label{f6}}

\figcaption[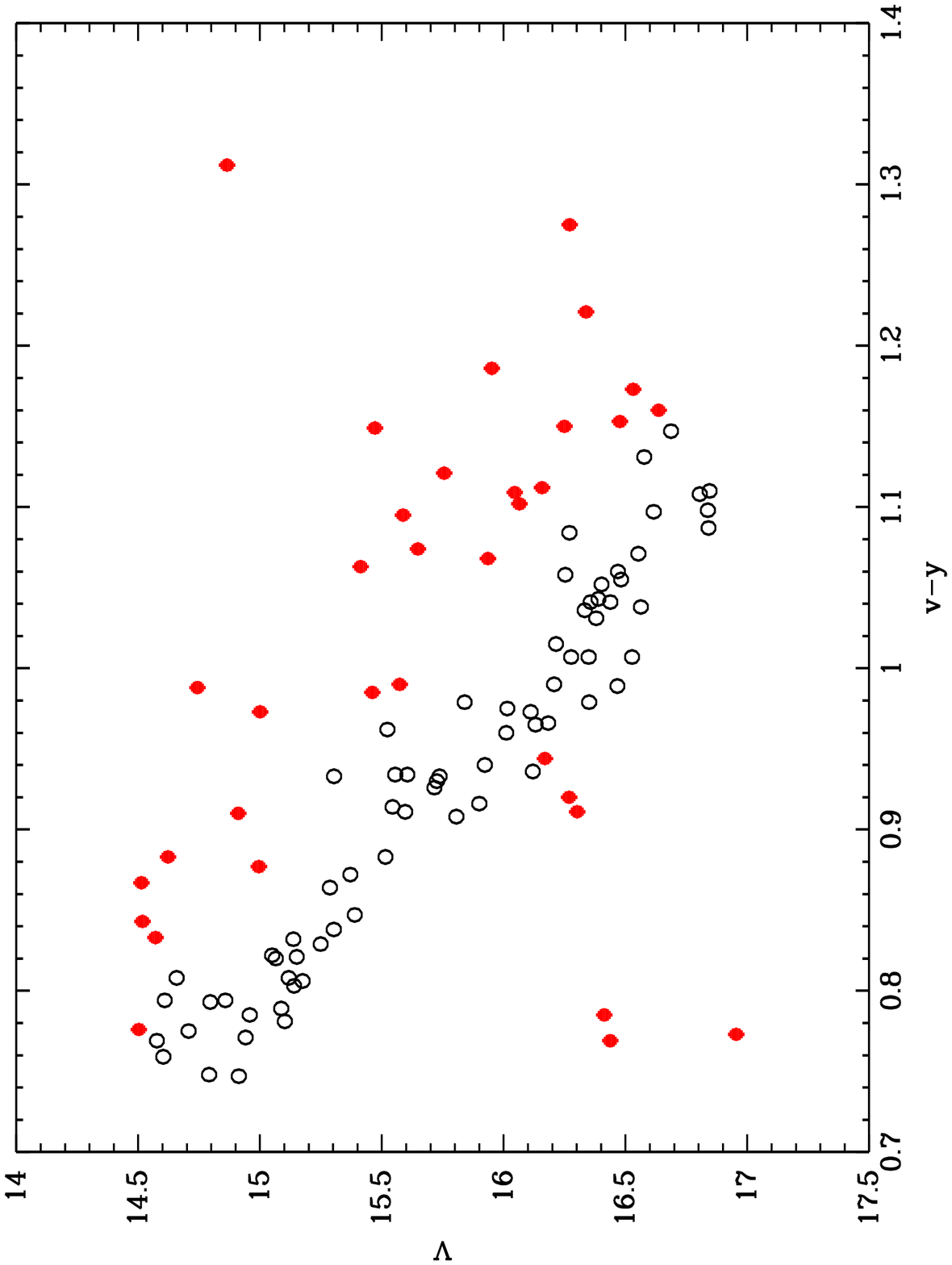]{$V, v-y$ CMD for the stars of Fig. 6. Symbols have the same meaning as in Fig. 6. \label{f7}}

\figcaption[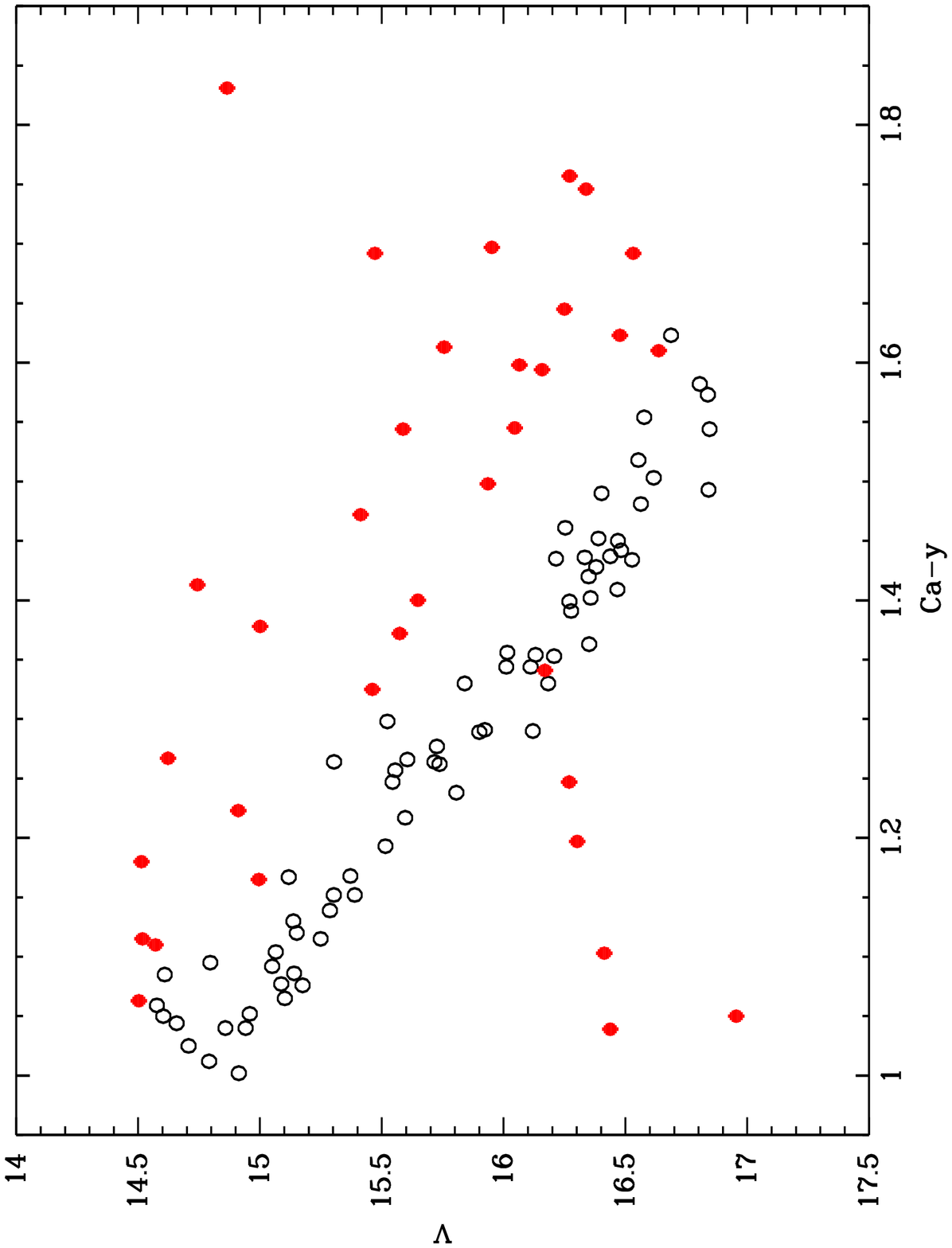]{$V, Ca-y$ CMD for the stars of Fig. 6. Symbols have the same meaning as in Fig. 6. \label{f8}}

\figcaption[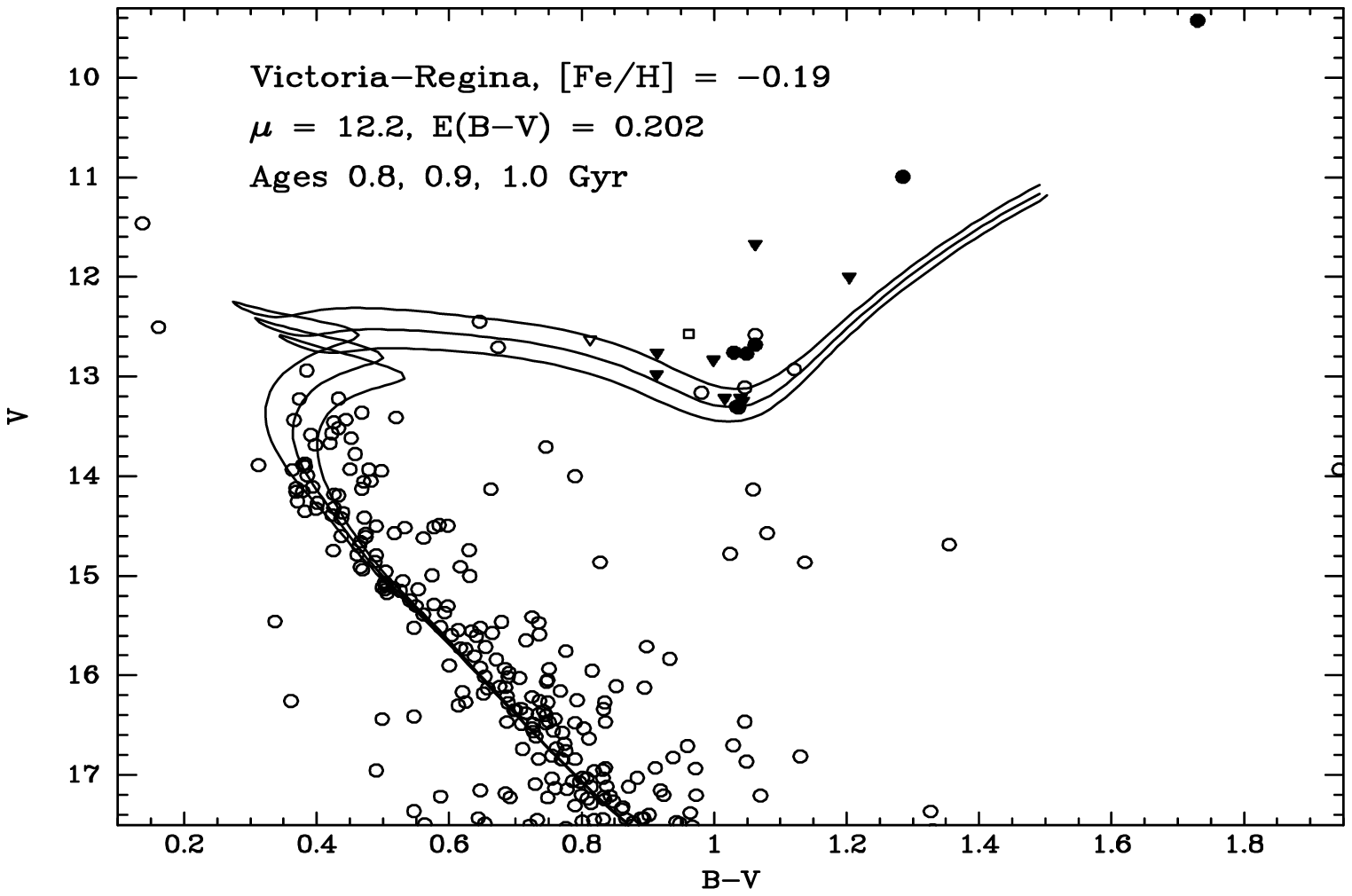]{CMD of Fig. 3, with a complete giant sample, converted to $BV$ and superposed on the [Fe/H] = --0.19, scaled-solar Victoria-Regina isochrones of \citet{vb06} with ages of 0.8, 0.9, and 1.0 Gyr, and adjusted for $E(B-V)$ = 0.202 and $(m-M)$ = 12.2. Filled circles are single-star, radial-velocity members, filled triangles are double-star members, the open square is a single-star probable member and the open triangle is a double-star, probable member. No membership information is available for the open circles. \label{f9}}

\figcaption[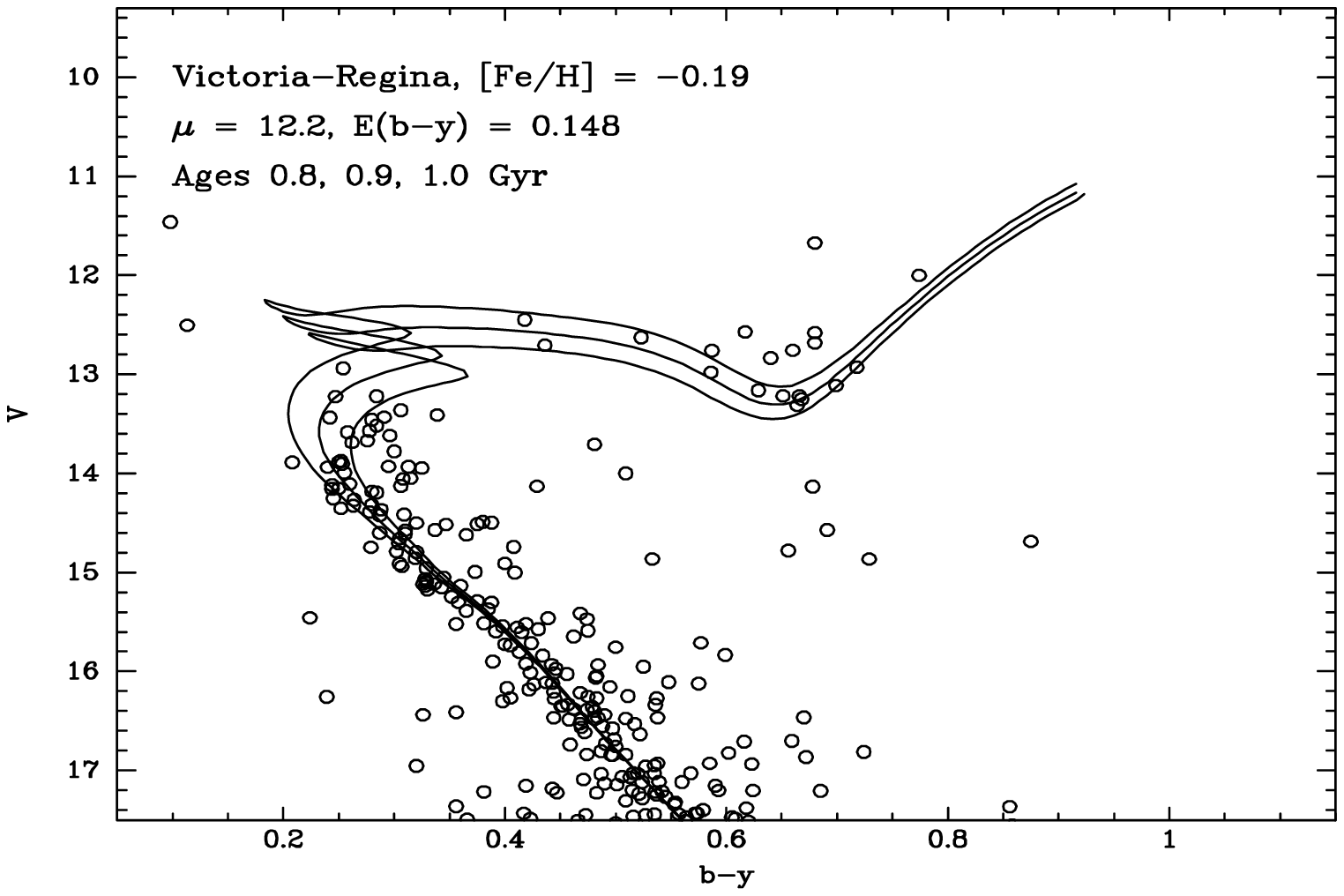]{Same as Fig. 9 for the $V,b-y$ Victoria-Regina isochrones. \label{f10}}

\figcaption[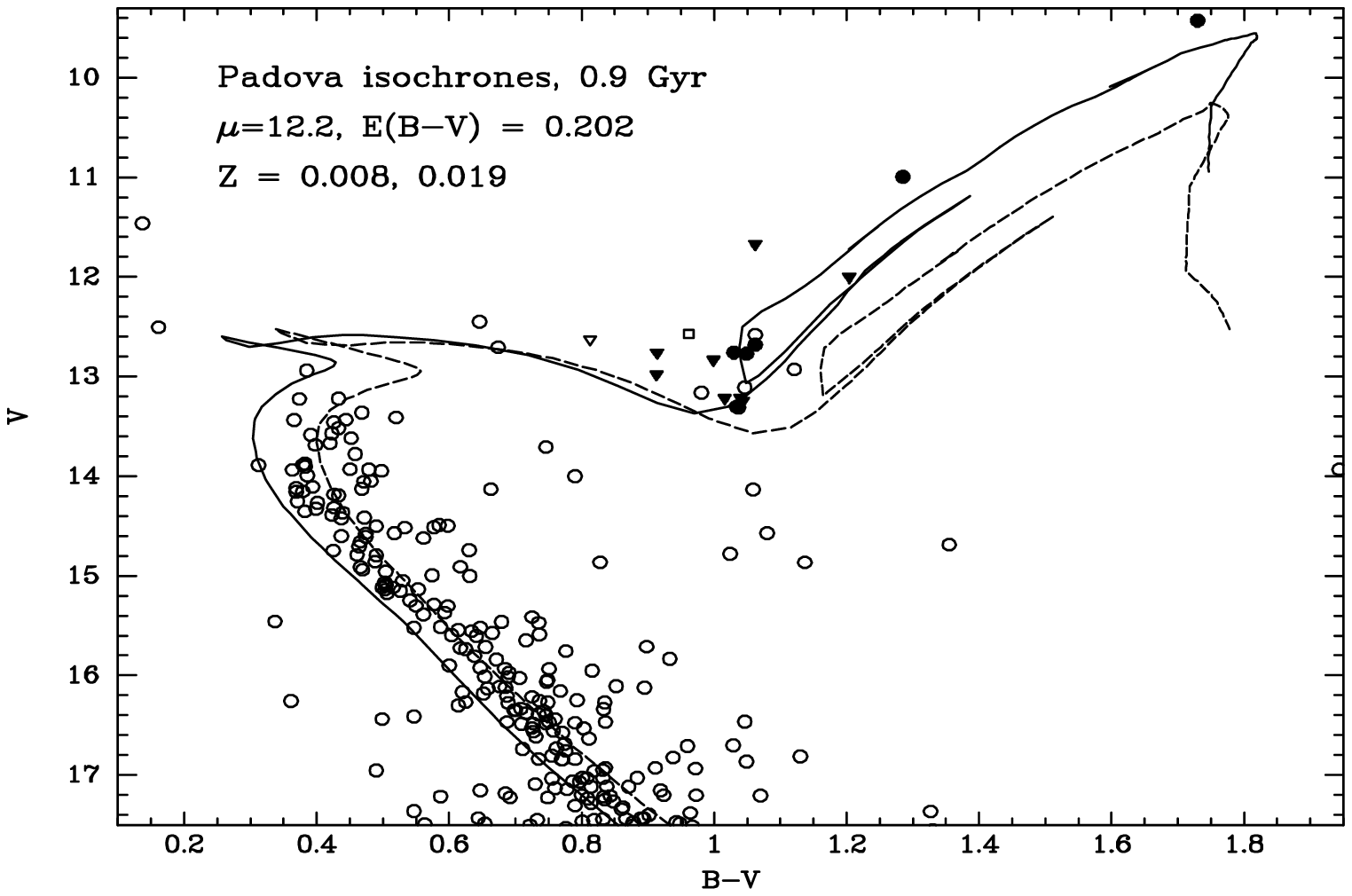]{Comparison between the data of Fig. 9 and the PAD overshoot isochrones. The two isochrones shown have the same age of 0.9 Gyr, but [Fe/H] = 0.00 (dashed curve) and [Fe/H] = --0.38 (solid curve), adjusted for $(m-M)$ = 12.2 and $E(B-V)$ = 0.20. \label{f11}}

\newpage
\plotone{f1.eps}
\newpage
\plotone{f2.eps}
\newpage
\plotone{f3.eps}
\newpage
\plotone{f4.eps}
\newpage
\plotone{f5.eps}
\newpage
\includegraphics[scale=0.53,angle=-90]{f6.eps}
\newpage
\includegraphics[scale=0.53,angle=-90]{f7.eps}
\clearpage
\includegraphics[scale=0.53,angle=-90]{f8.eps}
\plotone{f9.eps}
\newpage
\plotone{f10.eps}
\newpage
\plotone{f11.eps}
\clearpage
\thispagestyle{empty}
\begin{deluxetable}{rrrrrrrrrrrrrrrc}
\rotate
\tabletypesize\small
\tablenum{1}
\tablecolumns{16}
\tablewidth{0pc}
\tablecaption{Extended Str\"omgren Photometry in Melotte 71}
\tablehead{
\colhead{ID}     & 
\colhead{X}     & 
\colhead{Y}     & 
\colhead{$V$}     & 
\colhead{$b-y$}     & 
\colhead{$m_1$}     & 
\colhead{$c_1$}  &
\colhead{H$\beta$}     & 
\colhead{$hk$}     & 
\colhead{$\sigma_V$}     & 
\colhead{$\sigma_{by}$}     & 
\colhead{$\sigma_{m1}$}     & 
\colhead{$\sigma_{c1}$}  &
\colhead{$\sigma_{\beta}$}     & 
\colhead{$\sigma_{hk}$}     & 
\colhead{$N(ybvuwnCa)$}    }
\startdata
    83& -93.84& -66.06&  9.427& 1.126& 0.798& 0.134& 2.548 &1.791& 0.003&0.005&0.008&0.009&0.006&0.009&  6,8,7,9,8,12,5 \\
  2001& -93.88&-147.16&  9.671& 0.379&\nodata&\nodata& 2.586 &\nodata& 0.005&0.007&\nodata&\nodata&0.006&\nodata&  4,2,0,0,5,10,0 \\
    60& -66.41& -72.47& 10.293& 0.614& 0.431& 0.308& 2.536 &1.018& 0.003&0.005&0.008&0.010&0.005&0.008&  7,7,7,6,9,10,5 \\
    98&-135.42&  16.64& 10.694& 1.107& 0.879& 0.147& 2.535 &1.948& 0.003&0.005&0.007&0.010&0.003&0.007&  8,8,5,9,11,11,5 \\
    61& -63.53& -73.55& 10.949& 0.224& 0.168& 0.841& 2.724 &0.438& 0.003&0.005&0.008&0.009&0.004&0.008&  7,9,7,6,11,11,5 \\
   180&  92.82&  92.65& 10.984& 1.264& 0.860& 0.154& 2.565 &1.872& 0.006&0.007&0.010&0.010&0.004&0.012&  6,6,5,8,9,12,5 \\
  2002& 127.30&  10.68& 10.995& 0.765& 0.629& 0.281& 2.545 &1.294& 0.004&0.005&0.012&0.014&0.003&0.007&  6,9,5,6,10,8,5 \\
    23&  47.95& -46.09& 10.995& 0.828& 0.579& 0.235& 2.552 &1.253& 0.004&0.006&0.008&0.008&0.004&0.008&  8,7,6,8,11,7,5 \\
   346&  19.37& -19.03& 11.464& 0.098& 0.084& 0.721& 2.734 &0.137& 0.003&0.005&0.007&0.007&0.007&0.007&  8,7,5,6,6,4,5 \\
    69& -43.88&-131.86& 11.551& 0.690& 0.485& 0.305& 2.546 &1.157& 0.003&0.005&0.009&0.012&0.004&0.008&  7,7,5,6,10,9,5 \\
      &       & & & & & & & & & & & & & & \\
   114& -67.75&  -1.18& 11.622& 1.270& 0.770& 0.283& 2.551 &1.729& 0.007&0.012&0.017&0.019&0.011&0.016&  7,7,7,9,11,9,5 \\
   107& -80.34& -22.00& 11.673& 0.680& 0.358& 0.438& 2.596 &0.822& 0.003&0.005&0.008&0.009&0.004&0.009&  7,8,7,8,11,12,5 \\
   250&-129.73& 117.52& 11.734& 0.347& 0.164& 0.406& 2.636 &0.565& 0.004&0.005&0.008&0.009&0.005&0.008&  4,6,3,3,8,7,5 \\
   185&  54.97&  60.86& 11.737& 0.728& 0.416& 0.342& 2.569 &1.053& 0.005&0.006&0.011&0.013&0.005&0.008&  3,7,5,8,9,5,5 \\
    62& -69.10& -81.77& 11.878& 0.098& 0.176& 1.030& 2.863 &0.382& 0.003&0.005&0.008&0.009&0.005&0.007&  7,9,7,9,10,12,5 \\
   244&-121.65&  45.28& 11.893& 0.332& 0.184& 0.508& 2.666 &0.577& 0.004&0.005&0.007&0.008&0.005&0.007&  8,8,4,7,11,10,5 \\
   127& -37.17& -19.22& 12.001& 0.774& 0.532& 0.244& 2.542 &1.202& 0.003&0.004&0.008&0.010&0.005&0.007&  7,8,6,9,11,12,5 \\
     1& 123.00&-107.57& 12.080& 0.324& 0.142& 0.584& 2.690 &0.517& 0.004&0.006&0.008&0.008&0.004&0.009&  6,7,2,6,5,11,5 \\
   231& -50.63& 124.83& 12.237& 0.115& 0.193& 0.958& 2.952 &0.443& 0.003&0.005&0.009&0.010&0.006&0.007&  4,6,4,3,8,5,5 \\
   350& -25.29& -11.49& 12.451& 0.418& 0.181& 0.757& 2.719 &0.437& 0.003&0.006&0.009&0.010&0.005&0.008&  7,8,5,7,11,7,5 \\
\enddata
\end{deluxetable}

\end{document}